\documentstyle[aps,prb,epsfig,ecltree,epic,floats]{revtex}
\begin{document}

\twocolumn[\hsize\textwidth\columnwidth\hsize\csname@twocolumnfalse\endcsname

\title{Extrinsic models for the dielectric response
of $\mbox{CaCu}_{3}\mbox{Ti}_{4}\mbox{O}_{12}$}
\author{Morrel H. Cohen, J. B. Neaton, Lixin He, and David Vanderbilt}
\address{Department of Physics and Astronomy, Rutgers University,\\
Piscataway, New Jersey 08855-0849}
\date{April 8, 2003}

\draft
\maketitle

\begin{abstract}

The large, temperature-independent, low-frequency dielectric constant
recently observed in single-crystal CaCu$_3$Ti$_4$O$_{12}$ is most
plausibly interpreted as arising from spatial inhomogenities of its
local dielectric response.  Probable sources of inhomogeneity are the
various domain boundaries endemic in such materials: twin, Ca-ordering,
and antiphase boundaries.  The material in and neighboring such boundaries
can be insulating or conducting. We construct a decision tree for the resulting
six possible morphologies, and derive or present expressions for the 
dielectric constant for models of each morphology.  We conclude that all
six morphologies can yield dielectric behavior consistent with
observations and suggest further experiments to distinguish among them.

\end{abstract}


\narrowtext

]

\narrowtext


\section{INTRODUCTION}
\label{intro}  

The low-frequency limit $\epsilon_0$ of the dielectric
constant $\epsilon(\omega)$ of CaCu$_3$Ti$_4$O$_{12}$ (CCTO) 
single-crystals can be as much as 10$^3$ times
larger than that expected by extrapolating from
electron and infrared phonon contributions,
approaching 10$^5$ and remaining constant over a
broad temperature range.\cite{sub,ram,homes}  With a decrease in 
temperature or increase in frequency, $\epsilon$ falls off, 
displaying Debye-like relaxation behavior with an activated
relaxation rate.\cite{ram,homes}  Such unexpected behavior is startling
scientifically and intriguing technologically and, accordingly, has
attracted much attention,\cite{ram2,ram3,si,he,he2,pickett} including
comparative studies involving its Cd counterpart, CdCu$_3$Ti$_4$O$_{12}$ (CdCTO), 
which apparently shows similar though less pronounced behavior in 
ceramic samples.\cite{sub,homes-cd}

The central question is now whether the large dielectric
response is intrinsic to a perfect crystal of CCTO or extrinsic in 
that it originates with defects, inhomogeneities, etc., in particular
samples.  Based on our first-principles calculations of the
structure, the phonon spectrum, and the dynamical effective
charges of both CCTO (Ref.~\onlinecite{he}) and 
CdCTO (Ref.~\onlinecite{he2}), together with the agreement of our results
with experiment\cite{homes,homes-cd} and an assessment of the
existing experimental facts,\cite{sub,ram,homes,ram2,ram3,homes-cd}
we have argued strongly against an intrinsic interpretation, 
as have others on empirical grounds.\cite{sub,sinclair,lunk}  
At issue now is the nature of the
extrinsic mechanism, which we address in the present paper.

Empirically, the single-crystal samples of CCTO are known to 
be to be highly twinned,\cite{sub} and the transport behavior
of these domains and their boundaries, as well as those of other such domains 
and boundaries (in both single-crystal and ceramic samples), 
could play a significant role in the 
observed low-frequency giant dielectric response,
as was noted in earlier work.\cite{sub,he,sinclair}
The basic idea is that the bulk of the material is either
conducting or nearly so, and that the conductivity of the entire
sample is only prevented either by a failure of the conducting
regions to percolate, or else by the presence of thin insulating
blocking layers at the surfaces or at internal domain boundaries.
The various morphologies associated with these possibilities
can all be consistent with an enormous enhancement of the static
dielectric constant and with a Debye-like frequency response, although
there are characteristic details of the dielectric response that may
help distinguish between them.

In this article, we identify and thoroughly analyze those morphologies 
consistent with existing experimental results for CCTO.
We begin in Sec.~\ref{morph} by listing six morphology classes,
conceived from general arguments, that are associated 
with internal and external boundaries probably present in
CCTO and may give rise to the observed dielectric response.
Internal domain boundaries present in CCTO could
be conducting or nonconducting, and if the former is the 
case, the bulk must be insulating for a large dielectric response 
to be possible; alternatively if the latter is the case, 
the bulk must be conducting.  Of course, given either instance, 
intrinsic or electrode-induced insulating (or blocking) layers 
must be present at the surface if the internal conducting regions 
percolate through the sample, since a finite dielectric response
is observed.  And, if they do not percolate, 
blocking layers may or may not be present. After sorting out the
possible morphological scenarios, the dielectric response
of blocked morphologies is discussed in Sec.~III.  We then
provide useful exact bounds on $\epsilon_0$ in
Section~IV,  and afterward analyze the two unblocked morphologies
in detail, completing our analysis.
The dielectric properties of a solid exhibiting
bulk conductivity with insulating
boundaries and no blocking are derived in Section~V; the
converse morphology is treated in Section~VI.
We conclude, in Section~VII, that any of these six
morphologies would yield dielectric responses consistent
with observations,\cite{sub,ram,homes} and that existing experiments
do not distinguish them. 
Accordingly we propose various experiments, including
the use of nanoscale probes to examine the local conductivity,
to discriminate among the various possibilities presented here.

We do not consider here the possibility that the extrinsic
behavior arises from point or line defects randomly but
homogeneously distributed throughout the bulk.  One such
possibility would be that oxygen vacancies, common to perovskite-like
oxides and which we denote V$_{\rm O}$, are responsible. This
has already been proposed by Ramirez {\it et al.},\cite{ram4}
and they point out that a V$_{\rm O}$ concentration of order
10$^{-3}$ would be required.  If the activation energies of 
(V$_{\rm O}$)$^{+}$ and (V$_{\rm O}$)$^{++}$--i.e., 
singly- and doubly-ionized vacancies--are comparable to those
in other perovskites\cite{ram,dms} and if the mobilities of
electrons in the conduction band are similarly comparable,\cite{ihr,wemple,bern}
a concentration of V$_{\rm O}$ of order 10$^{-3}$ would
give rise to a dc bulk resisitivity many orders of magnitude
smaller than that observed or inferred for these materials.\cite{}
Accordingly we have confined our attention to two-dimensional
defects. 
 
\section{Six morphologies}
\label{morph}  

Large quasistatic dielectric constants
$\epsilon_0$ can arise in macroscopic insulators containing 
conducting regions approaching a percolation threshold.  
While this may occur in principle for a wide range of
morphologies, we consider here those six which we perceive
as relevant to the good single-crystal CCTO samples possessing the largest
values of $\epsilon_0$.\cite{homes} Each morphology
can be decomposed into two or three broadly-defined regions:
(1) domains, (2) their associated boundaries, and (3) blocking
layers (if any) parallel to the electrodes.  Several domain 
types (and therefore boundaries) are possible. 
In addition to twin boundaries, a necessary by-product of
the observed twinning,\cite{sub,homes,homes-cd} two other categories
may exist, both associated with variant chemical ordering.  
The first arises from the fact that the Ca sublattice may occupy four possible A sites, 
thus making possible four differently-ordered domains and
their associated domain boundaries.  The second includes
antiphase domain boundaries of A--A or B--B type separating domains
in which the A atoms have changed registry with the B atoms within
the perovskite structure.
Thus, if spatial heterogeneity in the electrical response occurs in
CCTO single crystals, it may be plausibly and economically 
associated with an intersecting set of locally planar twin,
antiphase, and/or compositional-ordering domain boundaries.

While each of the boundary regions contain planar defects, their
composition may differ from that of the bulk 
through the Gibbs adsorption phenomenon; it follows that
their effective thickness may be substantial in the present context.
We can thus distinguish two possibilities. The domains themselves
can be conducting or insulating in the bulk. If the domains are
conducting, their boundaries must be blocking, and vice versa. 
Impurities or oxygen deficiencies could
give rise to the bulk conductivity. Alternatively, if the
impurities or vacancies segregate at or near the boundaries,
the bulk would be insulating and the boundaries conducting.

Conductivity may also arise at a boundary through the formation of
bands of interface states, which are localized at the boundaries and
extended along them, and which possess a
fundamental gap signficantly smaller than the intrinsic bulk gap.
This would allow the formation of shallow traps associated
with the boundary imperfections. Conductivity could originate with
thermally activated traps or with ordinary thermal excitation
across the interfacial energy gap. For example, consider Cu-Cu 
antiphase boundaries.  CCTO is a Mott-Hubbard insulator with an 
intrinsic gap of at least 1.5 eV (Ref.\onlinecite{homes,he}).
The highest valence band and lowest conduction band states are primarily
$d$-orbitals hybridized with the $p$-orbitals of the four nearest-neighbor
oxygens in each CuO$_4$ plaquette.\cite{he} The coupling between
neighboring plaquettes is weak, as evidenced by the small valence and conduction
bandwidths. Within a tight-binding model, the effective Hubbard $U$ is thus
substantially larger than the effective number of neighbors $z$
times the effective electron transfer integral $t$ ($zt$ is then a 
measure of the bandwidth within this model). 
However, locally, at a Cu-Cu antiphase boundary, $z$ (and therefore 
the bandwidth) increases, $U/zt$ decreases,
the Hubbard gap decreases, and the possibility of boundary conductivity arises.
An excess V$_{\rm O}$ concentration could also give rise to
boundary conductivity as discussed further in Sec.~VII.

We conclude that in the present state of our knowledge of CCTO, 
we must suppose that the conducting bulk (CB) and the 
conducting interface (CI) morphologies
are equally plausible. The conducting regions can be below the percolation
threshold, i.e., nonpercolating (NP), or above the percolating threshold (P).
If the latter (P), conduction must be blocked by a surface barrier layer (SB).
The barrier layer can be intrinsic to the surface
or be associated with the electrode-CCTO interface.
In the nonpercolating case, there may or may not be a blocking layer.

\begin{figure}
\begin{center}
   \epsfig{file=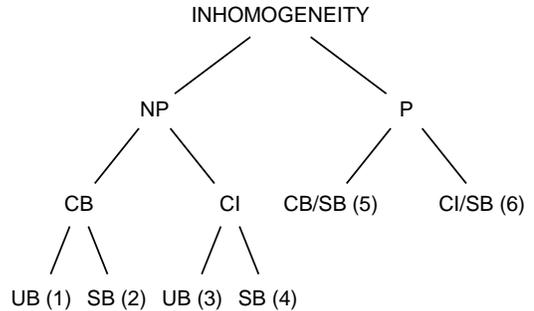,width=2.7in}
\end{center}
\caption{Morphological decision tree showing how the possible
morphological elements lead to six distinct morphologies
which are nonpercolating (NP), percolating (P),
conducting bulk (CB), conducting interfaces (CI),
unblocked (UB), or surface blocked (SB).
The numbers in parentheses correspond to those in Table~I.}
\label{fig:tree}
\end{figure}

These possibilities can be represented by a morphological
decision tree, illustrated
in Fig.~1. Six distinct morphologies result, which are listed in the Table~I
along with the acronyms by which we identify them in the remainder
of the paper.

\begin{table}
\squeezetable
\caption{Six possible morphologies that could produce the large
dielectric response.}
\label{tab:list}
\vskip 0.1cm
\begin{tabular}{lll}
 & MORPHOLOGY & ACRONYM \\
\hline
1 & nonpercolating, conducting bulk, unblocked & NP/CB/UB\\
2 & nonpercolating, conducting bulk, surface blocked & NP/CB/SB\\
3 & nonperc., conducting interfaces, unblocked & NP/CI/UB\\
4 & nonperc., conducting interfaces, surface blocked & NP/CI/SB\\
5 & percolating, conducting bulk, surface blocked & P/CB/SB\\
6 & percolating, conducting interfaces, surface blocked & P/CI/SB\\
\end{tabular}
\end{table}
%

\section{Blocked morphologies}
\label{analysis}  

We suppose that the CCTO samples have plane-parallel boundaries normal to
their smallest dimension, forming good planar electrode-sample
interfaces. The effective complex dielectric constant $\epsilon^*$
for all blocked cases is then simply
\begin{equation}
{1 \over \epsilon^*} = {f_{\rm L}\over\epsilon_{\rm L}} + 
{{1 - f_{\rm L}}\over\epsilon^*_{\rm M}} \; ,
\label{m1}
\end{equation}
where $f_{\rm L}$ is the volume fraction of the two blocking layers
which is equal to the sum of their relative thicknesses, 
$\epsilon_{\rm L}$ their dielectric constant (assumed to be entirely real), 
and $\epsilon^*_{\rm M}$ the complex macroscopic dielectric constant of the material.
We now focus on two possible sample morphologies, both containing conducting
regions, which lead to a large Debye-like low-frequency dielectric 
response; in one the conducting regions percolate, 
and in the other they do not.

\subsection{Dielectric response of
morphologies with percolating conducting regions}

For morphologies in which the bulk material contains 
percolating (P) conducting regions, the frequency-dependent dielectric constant
$\epsilon^*_{\rm M}$ can be expressed as 
\begin{equation}
\epsilon^*_{\rm M} = \epsilon_{\rm M}(\omega) + 4\pi i\sigma(\omega)/\omega,
\label{m2}
\end{equation}
with $\epsilon^*_{\rm M}$ and $\sigma(\omega)$ real.
Inserting (\ref{m2}) into (\ref{m1}) yields the Debye form
\begin{equation}
\epsilon^* = \epsilon_\infty + {{\epsilon_0 - \epsilon_\infty} \over 
{1 - i \omega\tau(\omega)}},
\label{m5}
\end{equation}
where
\begin{equation}
\epsilon^{-1}_0 = {\epsilon^*}^{-1}(\omega\rightarrow 0) = {f_{\rm L}\over\epsilon_{\rm L}},
\label{m6}
\end{equation}
\begin{equation}
\epsilon^{-1}_\infty = {\epsilon^*}^{-1}(\omega\rightarrow\infty) =
{f_{\rm L}\over\epsilon_{\rm L}}
+ {{1 - f_{\rm L}}\over\epsilon_{\rm M}(\infty)},
\label{m7}
\end{equation}
and
\begin{equation}
\tau(\omega) = 
{\epsilon_0\over\epsilon_\infty} {\epsilon_{\rm M}(\omega)\over{4\pi\sigma(\omega)}}.
\label{m8}
\end{equation}
In (\ref{m7}), $\epsilon_{\rm M}(\infty)$ is a representative value
of $\epsilon_{\rm M}(\omega)$ for $\omega$ such that 
$\omega\tau(\omega) \gg 1$.

It follows from the above that if a percolating morphology actually
occurs, it must be well beyond the percolation threshold.  This
results from the fact that $\epsilon_{\rm M}(0)$ diverges at the
percolation threshold, so that sufficiently close to the threshold
one expects no significant difference between $\epsilon_0$ and
$\epsilon_\infty$; then $\epsilon_{\rm M}(\infty) > \epsilon_0$, in
direct contradiction to the observations
($\epsilon_\infty\ll\epsilon_0$).

Neglecting $\epsilon_\infty$, the asymptotic (high-frequency)
behavior of the real part of $\epsilon^*$, denoted here $\epsilon'$, is
\begin{equation}
\epsilon' \sim {1\over\epsilon_0} {[4\pi\sigma(\omega)]^2\over \omega^2},
\label{m9}
\end{equation}
thus falling off with $\omega$ more slowly than the pure Debye case of 
$\omega^{-2}$.  Such behavior is also consistent with the experimental 
results of Ref.~\onlinecite{homes}, and thus the two percolating blocked
cases (P/SB, entries 5 and 6 of Table~I), cannot be eliminated on
the basis of their frequency dependence.

Using Eq.~(\ref{m6}) with a value of $\epsilon_{\rm L} \sim 10^2$, 
an $\epsilon_0$ of order 10$^5$ can be achieved if we take 
the blocking layer volume fraction as $f_{\rm L} \sim 10^{-3}$. 
As typical samples have thicknesses of
the order of millimeters, the implied widths of the blocking layers
are of the order of microns. It is unlikely, though not impossible,
that barrier widths are actually that large. Moreover, values of $\epsilon_0$ as
low as 5$\times$10$^3$ have been observed, implying barrier widths
of the order of tens of microns, which would be improbably large.
And finally, {\it conducting} samples have been fabricated,\cite{raevski}
which is inconsistent with the existence of such an electrode barrier
and mitigates against the P/CB/SB and P/CI/SB cases.

In the event that a percolating morphology is demonstrated to be the most
relevant to CCTO after all, a simple experiment
would distinguish between intrinsic surface and electrode-induced 
blocking layers. Suppose an insulating film of relative
thickness $f_{\rm I}/2$ and dielectric constant $\epsilon_{\rm I}$
is fabricated or deposited between each electrode and the sample.
The low frequency effective dielectric constant of the composite
(blocking layers plus insulating film), $\epsilon^0_{\rm eff}$, is given by
\begin{equation}
{1\over\epsilon^0_{\rm eff}} = 
{f_{\rm I}\over\epsilon_{\rm I}} + {f_{\rm L}\over\epsilon_{\rm L}}.
\label{m10}
\end{equation}
An intrinsic surface blocking layer should be affected but not eliminated
by the additional insulator. On the other hand, the effects of
an electrode blocking layer (EB) should certainly be eliminated.  Thus, measuring
$1/\epsilon^0_{\rm eff}$ for values of $f_{\rm I}$ greater than 
the $f_{\rm L}$ for the blocking layer and then extrapolating
the resulting $1/\epsilon^0_{\rm eff}$ vs.~$f_{\rm I}$ towards zero 
$f_{\rm I}$ should yield a vanishing intercept for the EB case. 
A similar procedure would yield a finite intercept for the 
intrinsic case, according to (\ref{m10}).  This experiment can be carried out
by depositing an electrode directly on one face of a thin CCTO
sample.  On the other face, a wedge of insulator is deposited followed
by the deposition of a set of separated electrode strips parallel
to the wedge boundary.  A set of values of f$_{\rm I}$ in (10)
embracing zero is thus achieved with a single sample.

\subsection{Dielectric response of 
morphologies with nonpercolating conducting regions}

For sample morphologies with nonpercolating conducting regions (NP),
we use for $\epsilon^*_{\rm M}$ the generic form
\begin{equation}
\epsilon^*_{\rm M} = \epsilon^{\infty}_{\rm M} +
{{\epsilon^0_{\rm M} - \epsilon^{\infty}_{\rm M}}\over
{1 - i\omega\tau_{\rm M}(\omega)}}.
\label{m11}
\end{equation}
The origin of this form will be discussed in more detail for the first four cases
of Table~I in Secs.~V and VI. Inserting (\ref{m11})
into (\ref{m1}) again yields the Debye form (\ref{m5}) but now with
\begin{equation}
\epsilon^{-1}_0 = {f_{\rm L}\over\epsilon_{\rm L}}
+ {{1 - f_{\rm L}}\over\epsilon^0_{\rm M}},
\label{m12}
\end{equation}
\begin{equation}
\epsilon^{-1}_\infty = {f_{\rm L}\over\epsilon_{\rm L}}
+ {{1 - f_{\rm L}}\over\epsilon^{\infty}_{\rm M}},
\label{m13}
\end{equation}
and
\begin{equation}
\tau(\omega) = 
{\epsilon_0\over\epsilon_\infty} 
{\epsilon^{\infty}_{\rm M}\over{\epsilon^0_{\rm M}}} \tau_{\rm M}(\omega).
\label{m14}
\end{equation}
For a value of $\epsilon_0$ close to the observed value of approximately
10$^5$, it follows from Eq.~(\ref{m12}) that 
$f_{\rm L}/\epsilon_{\rm L}$ and 
$(1 - f_{\rm L})/\epsilon^0_{\rm M}$ must be each 
less than 10$^{-5}$. Since $\epsilon_{\rm L}$ should be
of order 10$^2$, the value of $\epsilon_{\rm b}$, it follows
that $f_{\rm L}$ should be 10$^{-3}$ or less and $\epsilon^0_{\rm M}$
10$^5$ or larger. For a value of $\epsilon_{\infty}$ of
approximately 10$^2$, it then follows from Eq.~(\ref{m13}) that
$\epsilon^{\infty}_{\rm M} = \epsilon_{\infty}$=10$^2$. These two
results imply, with (\ref{m14}), that $\tau(\omega)\ge \tau_{\rm M}(\omega)$.

Interposing insulating films, as in the percolating morphology in
Sec.~A above, should not affect the intrinsic surface barrier. This implies that
\begin{equation}
{1 \over \epsilon^0_{\rm eff}} = {f_{\rm I}\over\epsilon_{\rm I}} + 
{f_{\rm L}\over\epsilon_{\rm L}} + 
{{1 - f_{\rm I} - f_{\rm L}}\over\epsilon^0_{\rm M}}
\rightarrow {1 \over \epsilon^0}
\label{m15}
\end{equation}
as $f_{\rm I}\rightarrow 0$. In the electrode-induced case the barrier at the electrode is
eliminated and $f_{\rm L}\rightarrow 0$ so that
\begin{equation}
{1 \over \epsilon^0_{\rm eff}} = {f_{\rm I}\over\epsilon_{\rm I}} + 
{{1 - f_{\rm I}}\over\epsilon^0_{\rm M}}
\rightarrow {1 \over \epsilon^0_{\rm M}} \ge
{1 \over \epsilon_0}.
\label{m16}
\end{equation}
In (\ref{m15}) and (\ref{m16}) the arrows signify an 
extrapolation to $f_{\rm I} = 0$ which is carried out
as described in Sec.~A above. Thus, if the extrapolated
value of $\epsilon^0_{\rm eff}$ exceeds $\epsilon_0$,
one learns from this experiment that the sample is blocked
by an electrode-induced layer and
that $\epsilon^0_{\rm M}$ exceeds $\epsilon_0$.
If $\epsilon^0_{\rm eff}$ does not exceed $\epsilon_0$ 
when $f_{\rm I}\rightarrow 0$, the experiment 
does not distinguish between the two possible sources of 
surface blocking.

\section{Exact bounds on the static dielectric constant}

Before discussing the nonpercolating unblocked cases (NP/CB/UB
and NP/CI/UB) and, equivalently, the origin of
expression (\ref{m11}) for $\epsilon^*_{\rm M}$
in the nonpercolating blocked cases of Sec.~IIIB, it is
convenient to have exact bounds on the total {\it static} dielectric 
constant $\epsilon_0$.

We make the simplification that the local dielectric and
conductivity tensors are isotropic.  Suppose an inhomogeneous
material has a set $\{\epsilon_i\}$ of different local
static dielectric constant with a corresponding set of volume fractions
$\{f_i\}$.  The static {\it macroscopic} dielectric constant $\epsilon_0$
then lies between the exact Wiener bounds\cite{torq}
\begin{equation}
\left[ \sum_i f_i\,{1\over\epsilon_i}\right]^{-1} \le \epsilon \le
\left[ \sum_i f_i\,\epsilon_i\right] \;,
\label{m17}
\end{equation}
the lower bound an equality for all inhomogeneities bounded by
planes parallel to the electrode planes (constant electric displacement 
$\bf D$) and the upper bound an equality for all inhomogeneities 
bounded by planes perpendicular to the electrode planes (constant 
electric field $\bf E$). When applied to the unblocked morphologies, 
these bounds are particularly illuminating, as we show below.

\section{Unblocked morphologies with insulating internal boundaries}
\label{blocking-boundaries}  

In the absence of blocking layers, 
we suppose that there is a bulk volume 
fraction $f_{\rm b}$ with finite conductivity $\sigma_{\rm b}$ and a
nonconducting internal interface (boundary)
volume fraction $f_i=1-f_{\rm b}$ with a finite dielectric
constant $\epsilon_{\rm i}$.  Since $\epsilon_{\rm b} \sim
4\pi i \sigma_{\rm b}/\omega\rightarrow\infty$ as
$\omega\rightarrow0$, Eq.~(\ref{m17}) simplifies to
\begin{equation}
\epsilon_{\rm i}/f_{\rm i}\le\epsilon\le\infty \;.
\label{m18}
\end{equation}
The upper bound pertains to systems above the percolation threshold.
Most interesting, however, is the lower bound; a small
volume fraction $f_{\rm i}\sim 10^{-3}$ associated with the
interfaces and interface dielectric constant of $\epsilon_{\rm i}$ comparable to
$\epsilon_{\rm b}$ together imply a lower bound comparable to the observed
$\epsilon_0$.

If all internal boundaries are parallel to the electrode
planes, the frequency-dependent dielectric constant becomes
\begin{equation}
\epsilon (\omega)= \epsilon_b +\frac{\epsilon_0 -\epsilon_b}
{1- i\omega \tau (\omega)} \; ,
\label{m19}
\end{equation}
\begin{equation}
\epsilon_0 = \epsilon_b /f_i\; ,
\label{m20}
\end{equation}
\begin{equation}
\tau(\omega)= (1-f_i) \epsilon_0 /4\pi \sigma_b (\omega) \; ;
\label{m21}
\end{equation}
with $f_{\rm i}$ the volume fraction of the internal blocking layers; with
the dielectric constant of the blocking layers assumed equal to
$\epsilon_{\rm b}$; and with $\sigma_{\rm b}(\omega)$ the bulk
conductivity within the domains.  A volume fraction $f_{\rm i}$ of $10^{-3}$
associated with the blocking domain boundaries would account for the
large observed value of $\epsilon_0$.\cite{homes}  Raevski {\it et al.}\cite{raevski}
have obtained dielectric responses similar to those of CCTO in the
nonferroelectric perovskite ceramics AFe$_{1/2}$B$_{1/2}$O$_3$ with A
being Ba, Sr, or Ca, and B being Nb, Ta, or Sb. They interpret their
results as indicating that a Maxwell-Wagner mechanism\cite{comment} is responsible
for the large permittivity and its temperature dependence, and they
use a simple layered structure to illustrate their argument empirically.
The layered geometry is certainly too artificial, however,
as in reality CCTO is likely to possess a more complicated (and
strongly nonplanar) arrangement of domains and boundaries.

A less unrealistic morphology would be spherical conducting domains
surrounded by spherical shells of insulator, with $\epsilon(\omega)$
derived within the effective medium approximation.\cite{torq3}  The
result is
\begin{equation}
\epsilon (\omega) = \frac{3 \epsilon_c(\omega) -2(\epsilon_c(\omega)
-\epsilon_b)f_i}{3 \epsilon_b +(\epsilon_c(\omega)
-\epsilon_b)f_i} \epsilon_b \; ,
\label{m22}
\end{equation}
where
\begin{equation}
\epsilon_c(\omega) = \epsilon_b+4\pi i\sigma_b(\omega)/\omega
\label{m23}
\end{equation}
is the complex dielectric constant of the conducting domains, and we
have set $\epsilon_{\rm i}=\epsilon_{\rm b}$ for simplicity.  This
can also be rewritten in the Debye form (\ref{m19}) with
\begin{equation}
\epsilon_0 = 3(1-f_i) \epsilon_b /f_i \; ,
\label{m24}
\end{equation}
\begin{equation}
\tau(\omega)=\epsilon_0 /(1-f_i)4\pi \sigma_b(\omega) \; .
\label{m25}
\end{equation}
Note that $\epsilon_0$ is increased above the lower bound in
(\ref{m18}) by a factor of $3(1-f_{\rm i})$ in this morphology,
according to the effective medium approximation.  In both
morphologies, $\epsilon_0$ increases monotonically as $f_{\rm i}$
decreases, a feature common to all nonpercolative morphologies.

The above simple treatments lead to deviations from Debye
relaxation only through the frequency dependence of $\sigma_{\rm
b}(\omega)$, which may or may not be significant for these materials
given our lack of understanding of the conduction mechanisms.  In
particular, randomness in the morphology can lead to such deviations,
particularly as close to the percolation threshold as the samples
appear to be, with $f_{\rm i}\sim$3$\times$10$^{-3}$ for 
$\epsilon_{\rm b}\sim$10$^2$ and $\epsilon_0\sim$10$^5$.
Our careful examination of the observed frequency
dependence of $\epsilon'$ and $\epsilon''$ shows that the expected
deviations from Debye-like behavior are indeed present.\cite{homes}  This case of
conducting domains enveloped by insulating domain boundaries is
identical to that of cermets with a low volume fraction of ceramic
or of varistors. Since these have been exhaustively studied,
we need comment no further on this case. Note that Eqs.~(\ref{m19}),
(\ref{m24}), and (\ref{m25}) provide a basis for the generic
form of $\epsilon^*_{\rm M}$, Eq.(\ref{m11}), introduced without
justification in Sec.~IIIB.

\section{Unblocked mophologies with conducting boundaries}
\label{noblocking}  

In this section, we present a model of a spatially inhomegenous material
whose bulk is insulating but whose internal boundary or interface regions 
are conducting. The model is developed from the
premise that the conducting boundaries are randomly parallel to one of three 
orthogonal cubic planes; as a simplification, we initially treat the boundaries as disjoint 
disks which we further simplify to oblate ellipsoids, identical except for their 
random orientation, with principal axes $a = b \gg c$ and occupying 
volume fraction $\phi$. For such a model to apply to CCTO, continuous conducting
pathways through the sample formed from boundaries are
forbidden, and so, while the conducting boundaries 
are allowed to intersect, we force the sample to be
in a nonpercolating regime.
The complex dielectric constant $\epsilon^*$ of 
our simple morphogenetic model system, derived below via an effective 
medium approximation, agrees quantitatively
with experiments, reproducing both the magnitude and frequency-dependence
of the observed dielectric response in CCTO and related materials.

We begin by supposing that our model material contains a
density $N$ of conducting boundary segments, which for
simplicity we take to be in the form of
disjoint thin discs of radius $a'$ and thickness $2c'$ randomly
distributed in the three cubic planes with volume fraction
\begin{equation}
\phi=2\pi N{a'}{^2}c'.
\label{a1}
\end{equation}
Since the overall symmetry is still cubic, we note that
$\epsilon$ is a scalar, and thus the response can be 
computed by assuming an electric field and polarization 
parallel to a single cubic axis. We now derive $\epsilon^*$ 
via an effective-medium approximation.
Suppose that each disc is at the center of a sphere of radius $b$
of the bulk material with complex susceptibility $\chi_{\rm b}$, 
where
\begin{equation}
\epsilon_{\rm b}=1+4\pi\chi_{\rm b},
\label{a2}
\end{equation}
outside of which is the effective medium.  The volume
fraction within the sphere of bulk material then becomes
\begin{equation}
\phi={3\over 2}{{a'}{^2}c'\over b^3} \;,
\label{a3}
\end{equation}
which leads to an important condition:
for the conducting discs to contain
the spheres completely and therefore be below the percolation threshold,
we must have ${2\over 3} {a'\over c'} \phi < 1$ 
in order to make contact with experiments.

\subsection{Bulk polarization}

The total macroscopic polarization contains contributions from the bulk-like
insulating domains, and also their associated internal 
conducting interfaces.  We derive each contribution in turn.
The polarization in the bulk material, $P_{\rm b}$, is related to the local
field within the bulk, $E_{\rm b}^{\rm loc}$, as 
\begin{equation}
P_{\rm b}=\chi_{\rm b}E_{\rm b}^{\rm loc} \;;
\label{a5}
\end{equation}
the local field inside the bulk material can be expressed in terms
of local and macroscopic polarizations, and also the macroscopic field, 
as 
\begin{equation}
E_{\rm b}^{\rm loc}=E_{\rm mac}+{4\pi\over 3}(P_{\rm mac}-P_{\rm b})
\;,
\label{a6}
\end{equation}
where $E_{\rm mac}$ is the macroscopic field in the effective medium
and $P_{\rm mac}$ is given by
\begin{equation}
P_{\rm mac}= \chi^*_{\rm mac}\, E_{\rm mac}={\epsilon^* -1 \over 4\pi} E_{\rm mac}
\; ,
\label{a7}
\end{equation}
where
\begin{equation}
\epsilon^*=1+4\pi\chi^*_{\rm mac} \;.
\label{a8}
\end{equation}

To obtain $\epsilon^*$, we need an expression for the macroscopic
polarization, $P_{\rm mac}$, in terms of $E_{\rm mac}$. 
We begin by first obtaining the former, $P_{\rm b}$, in terms of 
$E_{\rm mac}$. Inserting (\ref{a7}) into (\ref{a6}) gives
\begin{equation}
E_{\rm b}^{\rm loc}= {\epsilon^* +2 \over 3} E_{\rm mac} - {4\pi \over 3}
P_{\rm b} \; ;
\label{a9}
\end{equation}
now, inserting (\ref{a9}) into (\ref{a5}), we obtain the desired result
\begin{equation}
P_{\rm b}={{1 \over 3} (\epsilon^* +2) \chi_{\rm b} \over 1+ 
{4 \pi \over 3}\chi_{\rm b}} E_{\rm mac} \; .
\label{a10}
\end{equation}

\subsection{Polarization of the conducting internal interfaces}

Deriving the conducting-interface contribution to the macroscopic polarization 
is slightly more involved, and, to simplify the discussion without loss of generality,  
we now replace conducting discs with ellipsoids of the same 
volume $4\pi a^2c/3=2\pi {a'}{^2}c'$ and the 
same eccentricity 
\begin{equation}
e =\sqrt{\left({a \over c}\right)^2 -1}=\sqrt{\left({a' \over c'}\right)^2 -1} >>1,
\label{a11}
\end{equation}
where $a=(3/2)^{1/3}a'$, so that the polarization and
electrostatic field inside remain uniform.  The volume fraction 
of ellipsoids to bulk material (Eq.~(\ref{a3})) then becomes
\begin{equation}
\phi = {a^2 c\over b^3}.
\label{a12}
\end{equation}
The local field $E_{\alpha i}^{\rm loc}$ inside the ellipsoid 
differs from $E_{\rm b}^{\rm loc}$ by the depolarizing field 
introduced by the surface charge arising from the change in
polarization across the the ellipsoid-bulk interface:
\begin{equation}
E_{\alpha i}^{\rm loc}= E_{\rm b}^{\rm loc}+L_\alpha (P_{\rm b} - P_{\alpha i}) \;,
\label{a13}
\end{equation}
where $P_{\alpha i}$ is the polarization inside, $L_\alpha$ is the
depolarization factor, and $\alpha$ specifies the cubic axess normal
to the boundary plane, i.e., normal to the semi-minor axis $c$, where
$\sum_\alpha L_\alpha=4\pi$.
Using the linear relation $P_{\alpha \rm i}=\chi^*_{\rm i}E_{\alpha i}^{\rm loc}$, 
and also Eq.~(\ref{a13}), we find
\begin{equation}
P_{\alpha \rm i} = {\chi^*_{\rm i}(1 + L_\alpha \chi_{\rm b})
\over{1 + L_\alpha \chi^*_{\rm i}}} E_{\rm b}^{\rm loc},
\label{a15}
\end{equation}
where
\begin{equation}
\chi^*_{\rm i} = \chi_{\rm b} + {i\sigma_{\rm i}\over \omega}
\label{a16}
\end{equation}
is the complex susceptibility of the boundary region, $\sigma_{\rm i}$
its conductivity, and the real part of $\chi^*_{\rm i}$ has been
taken as $\chi_{\rm b}$ for simplicity.
Using (\ref{a9}) and (\ref{a10}), 
$E_{\rm b}^{\rm loc}$ can be expressed in terms of $E_{\rm mac}$ as
\begin{equation}
E_{\rm b}^{\rm loc}= {{1 \over 3}(\epsilon^* +2) E_{\rm mac} \over 
1+{4\pi \over 3} \chi_{\rm b}} ;
\label{a17}
\end{equation}
and from (\ref{a15}) and (\ref{a17}), we relate
$P_{\alpha \rm i}$ and $E_{\rm mac}$, as desired:
\begin{equation}
P_{\alpha \rm i}={{1 \over 3}(\epsilon +2) {\chi^*_{\rm i}(1 + L_\alpha \chi_{\rm b})
\over{1 + L_\alpha \chi^*_{\rm i}}}
{E_{\rm mac} \over 1+{4\pi \over 3} \chi_{\rm b}}} \; .
\label{a18}
\end{equation}

We now impose the effective-medium self-consistency condition 
that $P_{\rm mac}$ is the volumetric average of $P_{\rm b}$, i.e.,
\begin{equation}
P_{\rm mac}= (1-\phi) P_{\rm b} + \phi P_i \; ,
\label{a20}
\end{equation}
where we have taken the orientational average 
$P_i = {1 \over 3} \sum_{\alpha} P_{\alpha i}$.
The above equation, with the values of $P_{\alpha i}$
and $P_{\rm b}$ derived above, expresses $P_{\rm mac}$
linearly in terms of $E_{\rm mac}$. 

\subsection{Dielectric response}

We now recast $\epsilon^*=1+4\pi P_{\rm mac}/E_{\rm mac}$ in the suggestive form
\begin{equation}
\epsilon^* = {\epsilon_{\rm b} + {8\pi \over 3} B_{\rm i}\phi 
\over 1-{4\pi \over 3} B_{\rm i}\phi} \; ,
\label{a21}
\end{equation}
where we define
\begin{equation}
B_{\rm i} = {1\over 3} \sum_{\alpha} {1\over L_\alpha} {1\over {1 - i\omega\tau_\alpha}}
\label{a22}
\end{equation}
and
\begin{equation}
\tau_\alpha = \left({1\over L_\alpha} + \chi_{\rm b}\right) {1\over\sigma_i}.
\label{a23}
\end{equation}
The quantities $B_{\rm i}$ can be thought of as inverse depolarization
factors, with an additional dynamical correction.
Equation (\ref{a21}) is a generalization of the
Clausius-Mossotti relation which displays the proper limiting
behavior, as follows.  First, $\epsilon$ reduces to $\epsilon_{\rm b}$ when
$\phi$ vanishes.  Second, it reduces to the Clausius-Mossotti
relation in standard form for an array of metallic spheres for which
$\chi_{\rm b}=0$, $B_{\rm i}\rightarrow 3/4\pi$ as $\omega\rightarrow 0$;
that is,
\begin{equation}
\epsilon = {1 +{8\pi \over 3} N\alpha \over 1 -{4\pi \over 3} N\alpha}.
\label{a24}
\end{equation}
where $\alpha=a^3$ is the polarizability for the case that $c=a$.
Third, when $\sigma_{\rm i}\rightarrow 0$, the material becomes
uniform and $\epsilon^*$ reduces to  $\epsilon_{\rm b}$.

In the present case, B$_{\rm i}$ is the sum of two terms and
can be written explicitly as
\begin{equation}
B_{\rm i} = 
{2\over 3} {1\over L_a} {1\over{1 - i\omega\tau_a}} +
{1\over 3} {1\over L_c} {1\over{1 - i\omega\tau_c}} =
B_a + B_c,
\label{m26}
\end{equation}
where $B_a$ and $B_c$ are associated with the semimajor axes $a$ and $c$, and
\begin{equation}
\tau_a = (\chi_{\rm b}+1/L_a)/\sigma_{\rm i}
\label{m27}
\end{equation}
and
\begin{equation}
\tau_c = (\chi_{\rm b}+1/L_c)/\sigma_{\rm i}.
\label{m27a}
\end{equation}
From Landau and Lifshitz,\cite{landau} we obtain
\begin{equation}
L_c = 4\pi {1+e^2 \over e^3} (e-\tan^{-1} e) \; ,
\label{m28}
\end{equation}
\begin{equation}
L_a={1\over 2} (4\pi -L_c) \; ,
\label{m29}
\end{equation}
with $e$ given by (\ref{a11}).  For the oblate ellipsoids,
$a/c$ and $e$ are large enough so that $L_c$ approaches $4\pi$ and
$L_a$ becomes small,
\begin{equation}
L_a = \pi^2 {c\over a}.
\label{m30}
\end{equation}
Inserting $L_c = 4\pi$ and (\ref{m30}) for L$_{\rm a}$ into 
(\ref{m26}) and (\ref{m27}) yields
\begin{equation}
B_a = {2\over{3\pi^2}} {a\over c} {1\over{1-i\omega\tau_a}}
\label{m31}
\end{equation}
and
\begin{equation}
B_c = {1\over{12\pi}} {1\over{1-i\omega\tau_c}},
\label{m31a}
\end{equation}
with 
\begin{equation}
\tau_a = {{{4\over\pi} {a\over c} + \epsilon_{\rm b} + 1}\over{4\pi\sigma_{\rm i}}}
\label{m32}
\end{equation}
and
\begin{equation}
\tau_{\rm b} = {\epsilon_{\rm b}\over{4\pi\sigma_{\rm i}}}.
\label{m32a}
\end{equation}
$B_a$ and $B_c$ assume their maximum magnitudes at $\omega=0$;
explicitly,
\begin{equation}
B_a(0) = {2\over{3\pi^2}} {a\over c}
\gg  
B_c(0) = {1\over{12\pi}}
\label{m33}
\end{equation}
provided $a/c \gg 1$. Since $\phi \ll 0$ and $\epsilon_{\rm b}\sim$~80-90 
(Refs.~\onlinecite{homes,he}), $(8\pi/3) B_c(0) \phi$
is negligible relative to $\epsilon_{\rm b}$ in the numerator of
$\epsilon^*$, (\ref{a21}), and $(4\pi/3) B_c(0) \phi$ 
is negligible compared with unity in the denominator of $\epsilon^*$.
From (\ref{m31}), the ratio of the relaxation times is
\begin{equation}
{\tau_a\over\tau_{\rm b}} = {4\over\pi} {a\over{\epsilon_{\rm b} c}} +
1 - {1\over\epsilon_{\rm b}}.
\label{m34}
\end{equation}
Under the suppositions that $a \gg c$ and $\phi\ll 1$, it can be shown
from (\ref{m31}), (\ref{m32}), and  (\ref{m34}) that B$_{\rm c}$ does
not contribute significantly to $\epsilon^*$ at any frequency,
independent of the relative magnitudes of $a/c$ and $\epsilon_{\rm b}$
in (\ref{m34}).

B$_{\rm c}$ can thus be safely neglected, and $\epsilon^*$ can
be put into the Debye form with $\epsilon_\infty = \epsilon_{\rm b}$ and
\begin{equation}
\epsilon_0 = {\epsilon_{\rm b} + {8\pi \over 3}N \bar{a}^3 \over
1- {4\pi \over 3}N \bar{a}^3}.
\label{m35}
\end{equation}
Remarkably, Eq.~(\ref{m35}) allows the dielectric constant to be interpreted
as that arising from a collection of metallic spheres of effective radius
\begin{equation}
\bar{a} = \left({8\over 9\pi}\right)^{1/3} a \cong {2\over 3} a
\label{m36}
\end{equation}
embedded in a medium of dielectric constant $\epsilon_{\rm b}$.

Equation (\ref{m35}) can be rewritten as
\begin{equation}
\epsilon_0 = {\epsilon_{\rm b}+2 \bar{f} \over 1-\bar{f}},
\label{m37}
\end{equation}
where
\begin{equation}
\bar{f}={4\pi\over 3} N \bar{a}^3
\label{m38}
\end{equation}
is the volume fraction of the regions of the sample within which the
electrostatic field is screened out by the $a$-oriented conducting
boundaries, an analogue of the Faraday cage effect.  
The Debye relaxation time is now given by
\begin{equation}
\tau = {\tau_{\rm a}\over{1-\bar{f}}}.
\label{m39}
\end{equation}
As for Eq.~(\ref{m25}) for the NP/CB/UB case, all
temperature and frequency dependence of $\tau$ arises from
the conductivity $\sigma_{\rm i}$.

There is a pseudo percolation threshold at $\bar{f}=1$ at which $\epsilon$ in
Eq.~(\ref{m37}) diverges when the field-free regions percolate.
However, this mean-field theory and the geometry of the model both
grossly overestimate the value of
the percolation threshold through neglect of
randomness.\cite{torq} 
This overestimated threshold
occurs at a value of $\phi$ about 1.5\% below the true percolation
threshold of our disk-in-a-sphere model when the disk and sphere
radii coincide, also overestimated.

\section{Discussion and conclusions}
\label{conc}  

We have constructed six possible morphologies for inhomogeneity
in the local dielectric response of CCTO by supposing the properties
of the internal boundary regions to be opposite from those
of the bulk. (If one conducts, the other is insulating,
and vice versa.)  We have pointed out that the large
dielectric constants result either when the conducting
regions almost percolate or when they percolate but
are blocked at the surface or electrode interface.
We have analyzed the expected dielectric properties
of all six morphologies in Secs.~III-VI, in some cases
explicitly and in others by model calculation.
Without exception, the dielectric properties of
each of the six morphologies can be consistent
with all reported observations.\cite{sub,ram,homes}
Our principal conclusion is that existing experiments
do not distinguish between grossly different
morphologies, e.g., conducting bulk inhibited 
from percolation by insulating internal boundaries,
and conducting internal boundaries embedded in
and disrupted by insulating bulk.  As discussed in 
Sec.~III, distinguishing between intrinsic and electrode-induced
surface barriers can be achieved
by studying the dielectric response as a function
of the thicknesses of insulating layers interposed
between the CCTO sample and electrodes.  However,
the central questions are about the morphology
of the conducting regions: are they bulk or are they
associated with specific internal 2D faults?
The nonlinear current-voltage characteristics of the 
two morpholgies should differ quantitatively.
In both cases there should be a threshold field
at which current rises rapidly with voltage,
going over to a superlinear voltage dependence.
In the conducting boundary morphology (case 3),
field concentration would occur at the breaks in the
current pathways along the boundaries; 
this would not occur if the bulk were conducting (case 2),
and thus its threshold voltage would be higher.

Additional dielectric measurements at low frequencies ($<$1~MHz)
and temperatures (T$<$100~K) may be able to distinguish
morphologies with conducting bulk (CB) from those with conducting
boundaries or interfaces (CI). According to Homes {\it et
al.},\cite{homes} the {\it intrinsic} bulk dielectric constant
$\epsilon_0$ is enhanced by about 75\% as the temperature is reduced to
10~K from room temperature, a surprising increase which can be linked
mainly to an increase in oscillator strength of one low-frequency IR-active
mode.\cite{homes} This temperature dependence
would carry over to the extrinsic giant dielectric response in CI cases,
since in these scenarios the giant response comes about from
an amplification of the bulk $\epsilon_0$.
In CB morphologies, however, the intrinisic boundary $\epsilon_0$
may or may not be different from that of the bulk, and would
not necessarily exhibit a temperature dependence.
Thus, if measurements of the giant response at low temperatures
(and lower frequencies) did not show the increase expected from
that of the bulk $\epsilon_0$, one could infer that the CI
morphologies do not occur and that in the CB morphologies
the boundary $\epsilon_0$ differs significantly in its temperature
dependence from that of the bulk.
In addition, we note in both the CCTO and CdCTO $\epsilon_0$ increases
dramatically at higher temperatures. This would follow
directly from a thermally-stimulated increase in conductivity
locally throughout the material, bringing it closer to the
perocolation threshold. However, such would be the case
for all morphologies and, accordingly, is not diagnostic.

Additionally, what is the mechanism responsible for
the conductivity? The observed activation energy of
the relaxation time $\tau$ is that of the conductivity.
The values obtained are consistent with extrinsic 
conductivity associated with shallow traps. It would
be of interest to inject charge at one interface 
and measure the distribution of arrival times at
the other to see whether anomalous transport occurs. 

Observations with probes
sensitive to the conductivity, structure, and composition
at the nanoscale would be of significant
experimental interest. In Sec.~II we discussed 
a possible mechanism by which conductivity might arise
in a Cu-Cu antiphase boundary. In (Ba,Sr)TiO$_3$
A-A and B-B antiphase boundaries, Naumov {\it et al.}\cite{naumov} have found
compressive stress via first-principles computations.
This could be relaxed by a locally increased
concentration of oxygen vacancies leading to 
activated n-type conduction in the boundaries,
as discussed in Sec.~I for vacancies in the bulk.
Measurements of the sign of the thermopower
of samples with measureable conductivities 
could therefore be interesting. However, one should
recognize that the conducting boundary morphologies
require fine tuning of the boundary volume fraction
to remain close but not exceed the percolation threshold.

Finally it is important to recognize that the mechanisms leading
to large dielectric constants can differ in polycrystalline
(ceramic) and single-crystal samples.  For the former,
a model of conducting grains and blocking grain boundaries
may be the most plausible since, if the reverse were true,
the grain boundaries would percolate and there would
be substantial conductivity. For the latter, we cannot
yet distinguish between conducting and insulating
internal boundaries. Tselev {\it et al.},\cite{tselev} however, have 
argued for the conducting bulk and insulating
internal boundary morphology, the conductivity
being associated with oxygen vacancies and the
insulating boundaries being deficient in oxygen
vacancies.

\acknowledgments

This work was supported by NSF Grant DMR-9981193
and by the Center for Piezoelectric Design (CPD)
under ONR grant N0001401-1-0365.
We would like to thank I. Naumov, K. M. Rabe, and A. Tselev
for providing results prior to publication.



\end{document}